\documentclass[onecolumn,secnumarabic,amssymb, nobibnotes, aps, prd,12]{revtex4-2}

\usepackage{siunitx}
\usepackage{amsmath}
\usepackage{dsfont}
\usepackage{cancel}
\usepackage{graphicx}%
\usepackage{multirow}%
\usepackage{amssymb,amsfonts}%
\usepackage{amsthm}%
\usepackage{mathrsfs}%
\usepackage{hyperref}
\usepackage{natbib}
\begin{document}
\title{Probing High-Speed Electron Behavior in Magnetized Plasma Using Intense Laser Pulses and Quantum Electrodynamics}%
\author{{B.S. Sharma$^1$,Garima Yadav$^2$\footnote{Present address: Department of Physics, Lords University Alwar-India},D.N.Gupta$^3$\footnote{Present Address: Department of Physics and Astrophysics, Delhi University, Delhi, India} and N.K.Jaiman$^4$\footnote {Present address: Department of Physics, University of Kota, Kota, India}}}%
\affiliation{$^1$,Department of Physics, School of Science, Lords University,Alwar-301028,India}
\email{bssharma@lordsuni.edu.in}
\affiliation{$^2$,Department of Physics, School of Science, Lords University,Alwar-301028,India}
\email{garimayadavqwe@gmail.com}
\affiliation{$^3$,Department of Physics and Astrophysics, Delhi University-110007}
\email{dngupta@physics.du.ac.in}
\affiliation{$^4$,Department of Physics, School of Science, University of Kota-324005, India}
\email{nkjaiman@rediffmail.com}
\begin{abstract}
In this study, we utilize intense laser pulses and advanced quantum mechanical frameworks to investigate the behavior of high-velocity electrons within magnetized plasma environments. The focus of our research is placed within the context of strong-field quantum electrodynamics (QED), an area that explores the interaction of intense electromagnetic fields with charged particles. We propose new theoretical solutions that accurately describe these interactions by incorporating the effects of both the relativistic mass increase of fast-moving electrons and the influence of the electromagnetic field (or "light itself") within the plasma medium.
Our approach diverges from traditional models by offering a more comprehensive treatment of the complex dynamics at play in these extreme conditions. Standard models of laser-plasma interactions often fail to capture the full spectrum of physical phenomena that arise when strong magnetic fields are present. In particular, the traditional assumptions about electron motion and radiation emission become less accurate, as the interplay between the laser, the plasma, and the magnetic field introduces additional layers of complexity.
The novel solutions we present contribute to a deeper understanding of how radiation is emitted and interacts within magnetized plasmas, an essential aspect for advancing the development of next-generation laser-plasma accelerators. Our results underscore the limitations of conventional models and highlight the need for refined theoretical frameworks that can more accurately describe the intricate behaviors of particles and fields in these highly nonlinear environments.
\end{abstract}
\date{December 2024}%
\maketitle
\tableofcontents
\section{Introduction}
The interaction of intense laser pulses with plasma is a cornerstone of modern research in plasma physics and high-energy particle acceleration. With the advent of next-generation laser systems, the ability to generate extremely high fields and achieve ultra-fast electron dynamics has opened new frontiers in understanding fundamental plasma behavior.

Magnetized plasmas, in particular, introduce complex interactions due to the presence of strong magnetic fields that influence the motion of charged particles. Traditional models in plasma physics often treat particles as point-like, assuming non-relativistic speeds, and they rely on simplified electromagnetic field interactions. However, in highly intense laser fields, electrons can reach relativistic speeds, causing significant departures from these classical approximations.

Quantum Electrodynamics (QED) becomes increasingly important in such extreme environments, where the collective effects of strong electromagnetic fields, relativistic electron dynamics, and quantum processes (such as vacuum polarization) must be considered. Strong-field QED, a specialized branch of quantum mechanics, is required to accurately describe the interactions in these extreme conditions, where both the fields and particles behave in non-linear, coupled ways.

Despite advances in theoretical and computational plasma physics, many challenges remain in understanding how electrons behave in magnetized plasmas under high-intensity laser pulses. In particular, the effects of altered particle mass due to relativistic speeds, as well as the subtle influence of light itself, are not fully captured by traditional models. This research aims to fill these gaps, providing a deeper understanding of how radiation impacts electron behavior in magnetized plasma, and guiding the development of more accurate models for future laser-plasma accelerators.\\
In the realm of strong-field quantum electrodynamics (QED), Volkov solutions offer a powerful tool for unraveling the quantum behavior of charged particles, like electrons, interacting with intense electromagnetic fields.
Volkov solutions serve as a cornerstone of strong-field QFT, providing approximate solutions to the Dirac equation and illuminating the intricate dance of charged particles immersed in extreme electromagnetic environments.\\
Volkov solutions are approximate solutions to the Dirac equation for electrons in the presence of an external electromagnetic field. These solutions allow physicists to study how electrons respond to strong electromagnetic fields, including phenomena like high energy photon emission and electron-positron pair creation in strong fields. These solutions consider electron motion's classical and quantum aspects in a strong field. They are fundamental theoretical physics tools for understanding particle behavior in extreme conditions.\\
With the rapid advancements of high-intensity laser technology, laser pulses with intensity $\ge 10^{23}~\si {W{cm^{-2}}}$ has been achieved~\cite{Yanovsky08}. In years to come, the technologists may be able to bring much more power-focused laser facilities that could have intensities $\text I_a\gg 10^{23}~\si{Wm^{-2}}$. Such highly intense laser interaction with matter will open a new dimension to frontier physics. The dynamics of an electron at such high intensities can not be explained using theories of classical electrodynamics. It requires a new regime of physics called strong field quantum electrodynamics (SF-QED). The dynamics of charged particles are quite different in the SF-QED. The Lorentz factor $(\gamma)$ at such high intensities gets modified to much higher values $\approx 860 ~ (\text I_a\lambda_{\mu m})^{1/2}$, where $\text I_a \approx 10^{23}{W/cm^2}$ is laser intensity in a vacuum and material medium for planar linearly polarized electromagnetic waves. The   corresponding  maximum electric E$_0$ and magnetic  B$_0$ fields of I$_a\approx 10^{23}\si{W/cm^2}$  are $ \sim$  $2.7\times 10^{15}
{\text I_a}^{1/2}\si{V.m^{-1}}$  and $9{ \text I_a^{1/2}\si GG}$ respectively~\cite{Schwinger51,Salamin06}.\\
In the presence of a magnetic field of Mega-Gauss(MG) order, the  EM field of the intense laser pulse pushes to the SF-QED regime and beyond. The strong magnetic field also introduces anisotropy which becomes more imperative when an electron gyro-frequency $\Omega_e$=e$B_{eff}$/m ( throughout our discussion, we consider $c=1$ and $\hbar= 1$ ), where  $B_{eff}$ is an effective magnetic field of laser magnetic field and the external magnetic field, is no longer disregarded as compared with frequencies of relativistic plasma and intense laser pulse.\\
The magnitude of electric field at   intensity $10^{23}\si{W/cm^2}$ is much less than  the Schwinger critical electric field $\mathcal{E}_{cr} =m_e^2/e=1\cdot 32\times 10^{18} {\si V/m}$ and magnetic field $\mathcal{B}_{cr}=4.41\times 10^{9}$T ( a field that accelerates an electron to its rest mass energy over a  Compton wavelength~\cite{Schwinger51} and  lies in  sub-supercritical field). However, an electric field is not a Lorentz invariant in the rest frame of an ultra-relativistic electron. However, it can be boosted near the critical field using the external magnetic field of $1\ge$ MG. It allows studies of the Physics of laser-plasma interactions at sub-supercritical fields to elaborate the future technology of particle accelerators.\\
In low-density plasma, the role of a strong magnetic field is purely classical. However, quantum effects start emerging on increasing the plasma density and become essential in describing the interaction of laser plasma when the size of an electron wavelength becomes close to inter-particle distance. As the strength of the field further increases relative to Schwinger required field, $\mathcal{B}_{cr} \approx 10^{13} \si{G}$ the corresponding de Broglie wavelength shrinks to an electron Compton wavelength, and relativistic quantum effects become prominent.
These results in tempering the relativistic electrons wave function and momentum exchange between laser photon and nonlaser photon. The new modified wave functions are assumed to differ from plane wave solutions of Dirac's relativistic equation in the presence of heavy ions responsible for Bremsstrahlung and other nonlinear scattering processes such as Breit-Wheeler process  etc. Thus, in the presence of EM  waves of high intensities, the structure of electrons wave-packet and ponderomotive force changes. Also, relativistic electrons gain significant changes in their optimum  energy value
~\cite{Piazza12, Keit11, Angioi18, Blackburn20}.\\
When a strong  EM field propagates in a dense plasma, Dirac's relativistic wave equation of an electron modifies, and the closed analytic form of the solutions also changes ~\cite{Schroeder04,Bell08,Erber66}. However, in a strong EM  field, the Lorentz factor of an electron oscillates  to a value that is considered as the strength parameter $(a_0\gg 100)$ of EM waves~\cite{ Lau03}
\begin{equation}
	a_0=\frac{eE_0\lambda_{\mu m}}{2\pi m_e}=8.4\times10^2( I_a\lambda_{\mu m_e})^{1/2}\label{eq:1}
\end{equation}
where $m_e$ is the mass of plasma electrons. For $  I_a\ge {10^{23}~\si{W cm^{-2}}}$, the value of $ \text a_0$  is sufficiently high, and the interaction will be highly relativistic and nonlinear. Suppose the acceleration of an electron in its rest frame is compatible with the Schwinger acceleration $\text a_s= \text m_e$. In that case, classical electrodynamics cannot explain the physics of the interaction mechanism. In addition, as we increase the intensity, $\text a_0$  also increases. It turns to control multi-photon interaction in QED scattering processes. The  relevant parameters that characterize the interaction of electrons and photons in a strong EM field are defined as
\begin{equation}
	\chi=\frac{|\mathcal{F}_{\mu\nu}p^{\nu}|}{m_e\mathcal{E}_{cr}}
	\label{eq:2}
\end{equation}
where $\mathcal{F}^{\mu\nu}$ is the EM field tensor and $p^{\mu}$ is the corresponding particle momentum.
For $\chi\approx 1$, we have defined the fields as sub-supercritical fields. The quantum effects in the sub-supercritical fields change the interaction physics in the dense plasma when a strong EM field is present in the background. It necessitates a new kind of analytic solution to the Dirac equation for relativistic equations in a strong magnetic field. Accordingly, we have re-investigated the solutions of the Dirac equation beyond the famous Volkov solutions\\
The exact solutions of the Dirac equation in a strong field in SF-QED have been a study to understand new dynamics of a relativistic electron in dense magnetized plasma ~\cite{Volkov35, Gupta67, Peter65, Mendonca11, Franklin75, Ritus85, Ritus64, Bergou80}. \\ However, in this case, the Volkov solutions cannot be used directly in the present situation due to new properties of dispersion relations. The reason is that in the presence of relativistic magnetized plasma, the laser photon acquires an effective mass defined as an electron plasma frequency. Including the effective mass of photons makes a significant difference compared to Volkov's vacuum theory. The new solutions could serve as a basis for the description of possible quantum features  of radiating electrons in strong laser fields close to Schwinger field
.\\
The interaction of an ultra-intense short laser pulse with strongly magnetized plasma in a sub-supercritical field regime can be considered fully nonperturbative QED formalism as it involves multiphoton absorption. The dominant processes in such a strong field are nonlinear Compton scattering and nonlinear Briet Wheeler scattering, where an electron absorbs multiphoton and emits nonlaser $\gamma$  photons. Subsequently, a $\gamma$ photon decays into electron-positron pairs.~\cite{Popov06,Sokolov09}.
These new solutions of the process may serve as a basis for understanding the possible quantum theory of radiating electrons in strong laser fields~\cite{Kapusta89, Gong19}.\\
In novel plasma wake-field accelerator concepts, the generated electron beams are usually unpolarized \cite{Esarey09}. Most studies of interactions of the highly intense laser-plasma effect of the electron did not consider a strong magnetic near to Schwinger critical field. Vieira  \cite{Vieira11}  observed the spin precession of polarized electrons injected into a wakefield accelerator without a strong external magnetic field. In this paper, we establish the spin-dependent polarization properties of electrons in deriving new analytical solutions of the Dirac equation in strong magnetized plasma. The new solutions appear to establish a violation of the Volkov results.\\
This paper pushes the boundaries of our understanding by uncovering novel solutions to the Dirac equation in a realm bathed in intense light and swirling magnetism. Here, laser pulses ignite a dramatic tango between sub-supercritical electromagnetic fields, electrostatic waves, and the intrinsic spin and helicity of relativistic electrons.
Beyond familiar Volkov solutions, our newly found equations provide deep insights into the strange world of strong-field quantum electrodynamics (SF-QED). We paint a detailed picture of how spin-polarized electrons behave under extreme conditions, revealing the enigmatic influence of their helicity. This nuanced understanding unveils the essence of their relativistic dance within the dense embrace of magnetized plasma.
Our theoretical framework is no mere abstraction. It offers a precise roadmap for navigating the physics of exotic laser-plasma interactions and the delicate fingerprints of radiation reaction at intensities reaching a colossal $10^{23}\si{ W/cm^2}$. Moreover, it accounts for the subtle alterations imposed by the plasma environment, incorporating the shifts in electron and photon masses as light and matter become intricately intertwined.\\
Section II lays the foundation, presenting the modified model that underpins our new solutions and sets the stage for the SF-QED regime.
Section III dives into the heart of the matter, dissecting the intricate interplay of intense magnetic fields, electron polarization states, and helicity within the framework of the Dirac equation.
Finally, Section IV draws the curtain on this exploration, summarizing the groundbreaking insights gleaned from our analysis.
\section{  Constructing Solutions to the Dirac Equation: Unraveling Relativistic Electron Dynamics in Plasma}
In classical electrodynamics, the interaction strength of laser fields in a plasma medium is usually described as a dimensionless amplitude of electromagnetic vector potential, $ \text a_0=\text eE_0/m_e\omega$, where  $\omega$ and $\lambda$ are oscillating EM field frequency and wavelength respectively and  $\text E_0$ is the amplitude of the electric field. It represents work done by the fields over a distance $\lambda/2\pi$ in units of rest mass energy $m_e$. Considering
electromagnetic field tensor for the pump laser as
\begin{equation}
	e\mathcal{F}_{\mu\nu}=m_ea_0\sum_{i} f'_{i}(\phi)({k_{\mu}^{i}}{\varepsilon}_{\nu}^{i}-{k_{\nu}}{\varepsilon}_{\mu}^{i}),
	\label{eq:3}
\end{equation}
where k is  four wave-vector,  $f'_{i}(\phi)$ is the derivative of phase $\phi={ k} \cdot{ x}$, and $\varepsilon_{1,2}$ are constant polarization vectors  satisfy the condition $\varepsilon_i^2=-1$.\\
The four-momentum   $p^{\mu}$ of plasma electrons  in terms of  vector potential ~$
e\mathcal{A}_{\mu} =m_ea_0\sum f_i(\phi) \varepsilon^i_{\mu}$   is defined as:
\begin{eqnarray}
	p^{\mu}(\phi)
	=p_0^{\mu}+e\mathcal{A}^{\mu}-\left(	\frac{e\mathcal{A}\cdot{p}}{k\cdot p_0}+\frac{e^2\mathcal{A}^2}{2k\cdot p_0}\right)k^{\mu}.
	\label{eq:4}
\end{eqnarray}
where $p_0$ and $p$ are initial and final momentum, respectively.\\
Here, the transnational symmetry guarantees that ${k}\cdot{p}_{0}={k} \cdot{p}$. 
The net EM  field is the sum of the electromagnetic field of a strong laser pulse, background magnetic field ($\approx 1MG)$, and field of nonlaser photons radiated by accelerated electrons either through curvature radiation or through inverse Compton scattering of soft X-ray photons. These energetic photons interact with magnetized plasma and modify the structure of the wakefield and ponderomotive force. Using the Furry picture\cite{Furry51}, we decompose the electromagnetic potential in three parts
\begin{equation}
	{A}_{\mu}=\mathcal{A}_{\mu}^{\text{class}}+\mathcal{A}_{\mu}^q+\mathcal{A}_{\mu}^{\text{ext}}
	\label{eq:5}
\end{equation}
where  $\mathcal{A}^{\text{class}}$  is  the vector potential with respect to the ground state of the system. The ground state contains the ambient plasma and laser under consideration. $\mathcal{A}_{\mu}^q$ is quantum part which sands for the emitted radiation.  $\mathcal{A}_{\mu}^{\text{ext}}$  represents the external field imposed on the plasma. 
and ${\mathcal{A}}^{\text{ext}}=\partial ^{\nu}\mathcal{F}_{ \nu\mu}=0$ is an external covariant fields. Further $\mathcal{F}_{\mu\nu}  =\partial _{\nu}{A}_{\mu}-\partial_{\mu}{A}_{\nu}$ is electromagnetic field tensor.\\ The effective four vector potential can be written as
\begin{equation}
	{A}_{\mu}=\mathcal{A}_{\mu}^{\text class}+\mathcal{A}_{\mu}^q 	\label{eq:6}
\end{equation}
Since the intensity of optical lasers achieved with the present laser technology is far from the order $\sim 10^{29}{\si Wcm^{-2}}$, corresponding to Schwinger required field $\mathcal{E}_{cr}$. Thus, the effect of pair production is ignored in our present work.\\
The field ${A}$ propagates in the Dirac field $\psi$ of plasma electrons, and if the Dirac adjoint field is $\bar{\psi}$,  the effective Lagrangian   density $\mathcal{L_{\text{eff}}}$ of the entire interaction process can be written as
\begin{eqnarray}
	\mathcal{L_{\text{eff}}}&=\bar{\psi}(x)\left(\iota \cancel{\partial}- e\cancel{{\mathcal{A}}}- e{A}-m_e\right)\psi(x)\\ \nonumber &-e\bar{\psi}\cancel{\mathcal{A}}\psi-\frac{1}{4}\mathcal{F}_{\mu\nu}\mathcal{F}^{\mu\nu}
	\label{eq:7}
\end{eqnarray}
where $\cancel{A}=\gamma^{\mu}\partial_{\mu},\cancel{A}=\gamma_{\mu}A^{\nu}$, $m_e$ is the   electromagnetic mass of the electron in the strong electromagnetic field. Here $-e\bar{\psi}(x){\mathcal{A^{\text{class}}}}\psi(x) $  couples to the fermions to external field and $-e\bar {\psi(x)}\cancel{\mathcal{A}}^{\text{class}}\psi(x) \approx e\mathcal{A}^{0} j$,where $ j=\bar{\psi}(x)\gamma^{\mu}\psi(x)$ and $\mathcal{A}^{0}$ is field amplitude.
Introducing a characteristic momentum scale $Q=\iota\bar{\psi(x)\cancel\partial}\psi(x)$. The perurbative expansion is meaningful if $e\mathcal{A}^0 Q\le 1$. Taking account of this  fact, in non-relativistic fermion configuration, we define the field expansion parameter $\xi=e\mathcal{A}^0/m$ . If $\xi\gg 1$ the background has to be consider non-perturbatively, which amounts to diagonalizing the modified kinetic term. Thus in strong field quantum electrodynamics , the Lagrangian density  reduce to the following form.
\begin{equation}
	\mathcal{L}_{eff}=-\frac{1}{4}\mathcal{F}_{\mu\nu}\mathcal{F}^{\mu\nu}+\bar{\psi}(x)[\iota\cancel{\partial}-e\cancel{{{A}}}-m_e]\psi(x) 	\label{eq:8}
\end{equation}  
The resulting theory is known as the Furry picture of strong field quantum electrodynamics.  It preserve the structure of perturbative QED in its particle content and interaction term, photons and  so called dressed electron positron states interact via the bare QED interaction vertex, but qualitative differ from its vacuum equivalent field.\\
In our present model, we have considered the contribution of the effective mass of accelerated photon in terms of plasma frequency at maximum interaction between the accelerated photon and relativistic electron. As a result of such interaction, photons acquire a new momentum ${\mathbf k'}$
\begin{equation}
	k^{'\mu}k'_{\mu}=(\omega'^2-\mathbf{k'}^2)= m_{\text ph}^{ 2}=\omega_p^2
	\label{eq:9}
\end{equation}
where $\omega_p=\sqrt{ e^2n_e/\epsilon_0 m_e}$ with $n_e$ being the electron plasma density, $m_e$ electron mass, $k^{\mu}$ is the four-vector momentum and $m_{\text ph}$  as effective mass of photon.\\
The dynamics of the electron in SF-QED regime, we consider the electric and magnetic fields associated to the electromagnetic wave as $\mathbf{E}=-\partial_t\mathbf{A}$ and  $\mathbf{B}=(\mathbf{k'}\times {\hat\epsilon)}\partial_t {\mathbf A}$, where $\hat \epsilon$ is the unit polarization vector in the z direction. Thus, the quadratic  form of the Dirac equation for such interaction between relativistic plasma electron  and accelerated photon can be written as \cite{Mendonca11}
\begin{eqnarray}
	\big[\partial^2+2\iota e({{A}}^{})\cdot \partial)-e^2{{{A}}^{'^{2}}}+(m_e{}^2-{A}^{02})\\ \nonumber+\frac{1}{2}(\partial{{A}}^{})^2+\iota e\cancel{\bold{k'}}\cdot{{A'}}^{}+\frac{1}{2}\sigma^{\mu\nu}\mathcal{F}_{\mu\nu}\big]\psi_s(x')=0
	\label{eq:10}
\end{eqnarray}
where  $\partial^2 =\partial_{\mu}\partial^{\mu}=\nabla^2-\partial_t^2$ , ${A'}=\partial_t{{A}}$  ,$A^2=A_{\mu}A^{\mu}$, $A\cdot \partial=A^{\mu}\partial_{\mu}$, $\bold k'=(k',\omega')$ ,$\partial=(\vec{\partial},-\iota\partial_t)$,${ A}=\gamma^{\mu}A_{\mu}$,where $\gamma^{\mu}$ is Dirac matrices and $A$ is radiation field in Lorentz gauge. Here  we  have used $\psi_s(x')$  to emphasize that it is a solution of the second order Dirac equation at a new interaction point $x'$ under a unitarity transformation $k'=\omega'$ in  the relativistic notation, where  we have  assumed  a non-minimal coupling of the spin states $\sigma^{\mu\nu}$ via the field tensor term
\begin{equation}
	\frac{1}{2}\sigma^{\mu\nu}\mathcal{F}_{\mu\nu}=-\hat{\Sigma}\cdot \hat{\mathbf \mathcal{H}}+\iota{\mathbf \alpha}\cdot{\mathbf E}
	\label{eq:11}
\end{equation}
where helicity projection operator ${ \hat\Sigma}=(\mathds{1}+\sigma_z)/2$, with $\mathds{1}$ as a $4\times 4$ unit operator and $\sigma_z$ spin of particle. The projection operator $\Sigma$  defines the helicity of the  relativistic  polarized electron in the strong field quantum electrodynamics and
\begin{equation}
	\alpha_i=
	\begin{pmatrix}
		0&{\mathbf \sigma_i}\\
		{\sigma_i}&0
	\end{pmatrix}
	\label{eq:12}
\end{equation}
where $\sigma$'s is $2\times 2$ Pauli's spin operator, ${\mathbf \mathcal{H}}$ is interaction Hamiltonian of the system, and ${\mathbf E}$ is the electric field in the direction of propagation of laser photon.\\
Thus, the Dirac equation in the non-minimal coupling of spin with the electromagnetic field  for the upper(electron) and lower(positron) spinor components $\psi_{s+}(x')$ and $\psi_{s-}(x')$ in Eq.(12) can be written as
\begin{eqnarray}
	\Big[\partial_{\mu}\partial^{\nu}+2\iota e({A}\cdot \partial)-e^2{A}+(m_e{^2}-{A}^{o2})\\ \nonumber \pm\iota e\cancel{\bold{k'}}\cdot{{A}}\pm\frac{1}{2}(\partial_{\mu}\partial^{\mu}{A^{\nu}})\mp \iota(\hat{\Sigma}){k'}\cdot{A}\Big ]
	\begin{Bmatrix}
		\psi_{s+}(x')\\
		\psi_{s-}(x')
	\end{Bmatrix}
	=0
	\label{eq:13}
\end{eqnarray}
where we have approximate the function $1/2 \sigma ^{\mu\nu}\mathcal{F}_{\mu\nu}\approx\iota \hat {\Sigma} \cdot{A}^{} {k'}$ \cite {Berestetski82}.Alternatively, the above equation can be written as 
\\ The laser frequency $\omega$ in  a frame of  accelerated electrons transformed into ~\cite{Stenflo07,Mamun94,Ritus85}
\begin{equation}
	\omega'=\omega\sqrt{1-n^2}=\omega \sqrt{\frac{2\omega^2_{p}}{\omega^2-\Omega^2_{ce}}}
	\label{eq:14}
\end{equation}
where $n$ is the refractive index of plasma medium, $\Omega_{ce}$ is the cyclotron frequency, and $\omega_{pe}$ is the electron plasma frequency. \\
In the present model, we have considered radiation emission through curvature whose frequency approaches the cyclotron frequency. Cyclotron frequency can be considered close to laser frequency for a dense magnetized plasma and thus can be expressed in terms of the laser frequency.\\
The  generalized form of the Dirac equation  at some new field point $\phi'$ can be written as
\begin{eqnarray}
	\Big[\partial_{\mu}\partial^{\mu}+2\iota e({A}^{\mu} \partial_{\mu})-e^2\gamma^{\mu}\gamma_{\nu}{A}^{\mu\nu{2}}+(m_e^{^2}-{A}^{0}{}^2)\\ \nonumber+\iota e\gamma^{\mu}{k'}_{\nu}\cdot{{A}^{\nu}}+\frac{1}{2}(\partial_{\mu}{A}^{\nu})^2+\iota k'_{\mu}{\hat \Sigma}  {A}^{\mu}\Big ]
	\psi_s(\phi')
	=0
	\label{eq:15}
\end{eqnarray}
For a plane  electromagnetic wave, the vector potential  at new field point $\phi'$  can be read as
\begin{equation}
	{A}(\phi')=\left(a(\phi')\epsilon e^{\iota\phi'}+a^{*}(\phi')\epsilon^{*}e^{-\iota \phi'}\right)
	\label{eq:16}
\end{equation}
where $\phi' =\cancel{k'}\cancel{x'}=k'.x'=(\vec{k'}\cdot\vec{x'}-\omega'(k')\tau)$, with $\tau=t-(\vec{k}\cdot\vec{x}/\omega'(k'))$, is slow varying  function such that $\cancel{k'}\cancel{x'}=\cancel{k}\cancel{x}$ over a  Compton wavelength,  $a(\phi)=a^{*}(\phi)$ is  an envelope that slowly goes to $0$ as $\phi\rightarrow \pm\infty$ and
$\omega'(k')=\left | k'\right | $. The polarization vector obeys the conditions: $\epsilon^2=\epsilon^{*}{}^2=0$ as well as $\epsilon\cdot\epsilon^{*}=-1/2$ and  ${A}(\phi')\cdot {A}(\phi') =-a^2(\phi')$.\\
For the effect of all external EM fields, we have considered the polarization states of spin~$( \sigma=\pm 1/2)$ at $\cancel k'\cancel x'\rightarrow \pm \infty$ with   ${A}_{\mu}(-\infty)=0$. \\ 
We seek a solution for Eq.~(15) in the form
\begin{equation}
	\psi_s(\phi')=e^{\pm\iota {p}{x'}}f(\phi')
	\label{eq:17}
\end{equation}
Here, $p=p^{\mu}$ is constant four-vector momentum which is the momentum of the electron in the limit $k\cdot x \rightarrow \pm \infty$  and $f(\phi')$ is some function of the field point $\phi'$.\\
To solve this differential equation, we substitute the solution $\psi_s(\phi')=e^{\pm p  x} f(\phi')$ into  Eq.~(15), and then simplify to determine the specific form of $f(\phi')
$ based on the coefficients in the equation. 
Thus  from Eq.~(16), we can write a second-order differential equation
\begin{eqnarray}
	\Big[m_{e}^2f''(\phi')-(2e{A_{\mu}} k'^{\mu}+k'_{\mu}p^{\mu})f'(\phi')\\ \nonumber-e^2{A}^0{}^2)f(\phi') +\iota e(\cancel{k'}{A})+\frac{1}{2}(\partial_{\mu}{A}^{\nu})^2-\iota (k'{\hat \Sigma} \cdot (A)f(\phi')\Big ]=0
	\label{eq:18}
\end{eqnarray}
For a plane wave traveling in a relativistic magnetized plasma, $m_e^2\ne0$. This is a second order differential equation and to solve the equation for $f(\phi')$, we have to solve the homogeneous part first:
\begin{eqnarray}
	\Big[m_{\text ph}{}^2f''(\phi')-2e({A}_{\mu}p^{\mu}+k'_{\mu} p^{\mu})f'(\phi')-(2e{A_{\mu}} k'^{\mu} -e^2{A}^0{}^2)f(\phi')\Big]=0\label{eq:19}
\end{eqnarray}

Assuming a solution of the form $f(\phi')=e^{\lambda\phi'}$ and substituting into the homogeneous equation, we get
\begin{eqnarray}
	\lambda=\frac{e(A_{\mu}k'^{\mu}+k'_{\mu}p^{\mu})\pm \sqrt{2e(A_{\mu}k'^{\mu}+k'_{\mu}p^{\mu})^2+2m_e^2(2eA\cdot k'+e^2A^{02})}}{m_{\text ph}^2e^{2}}\label{eq:20}
\end{eqnarray}
The general solution to the homogeneous equation is :
\begin{equation}
	f_{h}({\phi'})=C_1e^{\lambda_1\phi'}+C_2e^{\lambda_2\phi'}	\label{eq:21}
\end{equation}
where $\lambda_1$ and $\lambda_2$ are the roots found above.
The particular solution of inhomogeneous terms 
\begin{eqnarray}
	f_p(\phi')=-\frac{1}{2}(\partial_{\mu}{A}^{\nu})^2-\iota (k'{\hat \Sigma} \cdot (A)f(\phi')=0 	\label{eq:22}
\end{eqnarray}
The final solution  for $f(\phi') $is:
\begin{eqnarray}
	f({\phi'})=C_1e^{\lambda_1\phi'}+C_2e^{\lambda_2\phi'}-\frac{1}{2}(\partial_{\mu}{A}^{\nu})^2-\iota (k'{\hat \Sigma} \cdot (A))	\label{eq:23}
\end{eqnarray}
The value of constant $C_1+C_2$ can be determined using the boundary conditions as explained above. The final solution can be written as
\begin{eqnarray}
	f(\phi')\sim e^{\int_0^{\phi'}\frac{1}{\omega_p^2}(e(A_{\mu}k'^{\mu}+k'_{\mu}p^{\mu})d\phi'}\\ \nonumber\pm  e^{\int_0^{\phi'}\frac{1}{\omega_p^2}\sqrt{2e(A_{\mu}k'^{\mu}+k'_{\mu}p^{\mu})^2+2\omega_p^*{}^2(2eA\cdot k'+e^2A^{02})}d\phi'} \left[-\frac{1}{2}(\partial_{\mu}{A}^{\nu})^2-\iota (k'{\hat \Sigma} \cdot (A))\right]\frac{1}{\omega_p^2}u_{p,\sigma}(\phi')\label{eq:24}
\end{eqnarray}
where $u_{p,\sigma}(\phi')$ represents the spinor of an electron in the new polarized state at field point $\phi'$. Thus, the wave function is reduced to the following form
\begin{eqnarray}
	\psi_{p,\sigma}(x')\sim e^{\pm\iota px'-\int_0^{\phi'}\frac{1}{\omega_p^2}(e(A_{\mu}k'^{\mu}+k'_{\mu}p^{\mu})d\phi'}\\ \nonumber\pm  e^{\int_0^{\phi'}\frac{1}{\omega_p^2}\sqrt{2e(A_{\mu}k'^{\mu}+k'_{\mu}p^{\mu})^2+2\omega_p\label{key}{}^2(2eA\cdot k'+e^2A^{02})}d\phi'} \left[-\frac{1}{2}(\partial_{\mu}{A}^{\nu})^2-\iota (k'{\hat \Sigma} \cdot (A))\right]\frac{1}{\omega_p^2}u_{p,\sigma}(\phi')	\label{eq:25}
\end{eqnarray}
In Eq.~(25), the exponent term is somewhat similar to classical action. Thus, the solution of the Dirac Equation  under the said conditions can be written as
\begin{eqnarray}
	\psi'_{p,\sigma}(x')\sim \frac{1}{\sqrt{2\mathcal{E}_0}}e^{\iota S(\phi')} \left(-a^2k'^2 -\iota (k\hat{\Sigma} a)\right)\frac{\omega'2}{\omega_p^2}~u_{p,\sigma}(\phi')
	\label{eq:26}
\end{eqnarray}
In writing Eq.~(26),  we  have used Eqs.~(9) and Eq.~(16).
Here $1/\sqrt{2\mathcal{E}_0}$ is a normalized function  with $\mathcal{E}_0$ is an initial energy of particles such  that $\int \psi_{p,\sigma}^{\dagger}\psi_{p,\sigma}d\tau=1$, where $d\tau$ is volume element  in field space.\\
The action of magnetized plasma electrons in the strong electromagnetic plane wave  is given by
\begin{eqnarray}
	S(\phi')= -p'\cdot x'- \int_0^{\phi'}\frac{\omega'^2}{\omega_p^2}\Big(1+\frac{e {k'}{a}}{2k'\cdot p'}+\frac{2p'^2a^2}{k'^2p'^2}+\frac{ea.k'}{k'^2p'^2}\Big)d\phi'
	\label{eq:27}
\end{eqnarray}
Using explicit form of ${A}$,  Eq.~ (19) can   be rewritten as
\begin{eqnarray}
	m_{\text ph}^2f''(\phi')-2\iota ek'\cdot p'f'(\phi')+\Big(-2ea^{*}(p\cdot\epsilon^{\mu}\cdot \epsilon_{\nu}^*+\hat{\Sigma} e{k'}_{\mu}{\epsilon}^{\nu}(a\cdot p)e^{\iota\phi'}\nonumber \\
	-ae {k'}^{\mu}{\epsilon}_{\nu}^{*}e^{\iota\phi'}  -2e(p\cdot\epsilon_{\mu}^*\epsilon^{\nu})a^{*}e^{-\iota\phi'}+\iota ek'\cdot ae^{\iota\phi'}+\nonumber \\ \hat{\Sigma} k'\cdot a^{*}e^{-\iota\phi'}-e^2(a^2+a_0^2)\Big)f(\phi')=0
	\label{eq:28}
\end{eqnarray}
where $a_0=e{A}_0/m_e$ where  ${A}_0$ peak value of the vector potential at t~=~0 and $a=eA/m_e$.Further, here we consider $a$ and $a*$ as annihilation and creation operators satisfies the condition $[a_i,a_j*]=\delta_{ij}$. Eq.~(28) can be written in the following form
\begin{eqnarray}
	m_{\text ph}^2f''(\phi')-2\iota  k'\cdot p'+\Big(\chi e^{\iota\phi'} + \eta e^{-\iota\phi'}-2e^2a_0^2\Big)f(\phi')=0
	\label{eq:29}
\end{eqnarray}
where we assume $a\sim a_0$ and
\begin{eqnarray}
	\chi =e(a\cdot p'+\sigma {k'}^{\mu}{\epsilon}_{\nu}\cdot a
	-{k'}_{\mu}{\epsilon^*}^{\nu}\cdot a)\\ \label{eq:27}
	\eta= e(p'\cdot a+ k'\cdot a+\hat{\Sigma} k'\cdot a)
	\label{eq:30}
\end{eqnarray}
both $\chi$ and $\eta$ are related to four vector potentials and are written as $4\times 4$ matrix. The remaining terms are scalars.
Under Lorentz gauge $(k'\cdot a)=0$  and    $\chi^2=\eta^2=0$,
\begin{eqnarray}
	\chi\eta=-e^2a^2(\omega'^2+\hat{\Sigma}m_{\text ph}^2)\mathcal{M}_1,
	\label{eq:31}\\
	\eta\chi=-e^2a^2(\omega'^2+\hat{\Sigma}m_{\text ph}^2)\mathcal{M}_2
	\label{eq:32}
\end{eqnarray}
where $\mathcal{M}_1$ and $\mathcal{M}_2$ are diagonal matrices defined as
\begin{equation}
	\mathcal{M}_1=
	\begin{pmatrix}
		\hat{\Sigma}&0&0&0\\
		0&\Sigma&0&0\\
		0&0&\hat{-\Sigma}&0\\
		0&0&0&-\Sigma
	\end{pmatrix} \quad  \text{and}~~\mathcal{ M}_2=
	\begin{pmatrix}
		\Sigma&0&0&0\\
		0&\hat{-\Sigma}&0&0\\
		0&0&\Sigma&0\\
		0&0&0&\hat{-\Sigma}
	\end{pmatrix},
	\label{eq:33}
\end{equation}
The Eq.~(29) is similar to the Mathieu equation and can be deduced by considering function $f(\phi')$  a periodic function. According to Floquet's theorem
\begin{equation}
	f(\phi')=q(\phi') e^{\iota b \phi'}
	\label{eq:34}
\end{equation}
where
\begin{equation}
	q(\phi') =\sum\limits_{n=-\infty}^{\infty}\nu_ne^{\iota n\phi'},
	\label{eq:35}
\end{equation}
b and $\nu_n$ are constant bispinors, and n is an integer.
Using Eq.~(35) in Eq.~(29), we get
\begin{equation}
	m_{\text ph}^2q''(\phi')+\beta q'(\phi')
	+(\Delta+\chi e^{\iota\phi'}+\eta e^{-\iota\phi'})q(\phi')=0,
	\label{eq:36}
\end{equation}
where
\begin{equation}
	\beta=2\iota\left(b m_{\text ph}^2-e k'\cdot p'\right)
	\label{eq:37}
\end{equation}
\begin{equation}
	\Delta=m_{\text ph}^2b^2-2bk'\cdot p'-2e^2a^2.
	\label{eq:38}
\end{equation}
Using Eq.~(33) in Eq.~(34), we get
\begin{equation}
	\Xi_n\nu_n+\chi\nu_{n-1}+\eta\nu_{n+1}=0
	\label{eq:39}
\end{equation}
with
\begin{equation}
	\Xi_n=(m_{\text ph}^2n^2+\iota\beta n+\Delta)
	\label{eq:40}
\end{equation}
Substituting  Eq.~(38) and Eq.~(39) in Eq.~(41), we get
\begin{equation}
	\Xi_n=m_{\text ph}^2(n+b)^2-2k'\cdot p'(n+b)-2e^2a^2
	\label{eq:41}
\end{equation}
The vector coefficients in Eq.~(40) are matrix to get the finite set of matrix equations. We consider n  for finite values from -N to N with $N\gg 1$. Being a $4\times 4$ matrix, the total  dimensions of $\mathcal{M}$ are  $4\times (2N+1)$. Thus, the set of matrix equations can be described as
\begin{equation}
	\mathcal{M}\nu=0
	\label{eq:42}
\end{equation}
where
\begin{equation}
	\mathcal{M}=
	\begin{pmatrix}
		\hat{ \Sigma}	\Xi_{-N}I_4 & \chi&&&&0\\
		\eta &\hat{\Sigma}\Xi_{-N+1}I_4& \chi&&\\
		\vdots&\ddots&\ddots&\ddots &&\\
		&&&&\eta&\hat{\Sigma}\Xi_NI_4
	\end{pmatrix}
	\label{eq:43}
\end{equation}
with
\begin{equation}
	\nu=
	\begin{pmatrix}
		\nu_{-N}\\
		\vdots\\
		\nu_{N}
	\end{pmatrix}
	\label{eq:44}
\end{equation}
where $I_4=\mathds{1}$ is a $4\times 4$ matrix.
The solution of Eq.~(41)  determines any nonzero value of $b$  for which at least one coefficient of $\Xi_i$ is nonzero for $i\neq r$ and $i=r$ where r is any unknown fixed integer value. Out of several possibilities, we choose $r=0$ for which $\nu$ takes a nonzero value $\nu_0$.\\
Thus, the transformation equation corresponding to the first row can be obtained by diagonalizing Eq.~(38). The equation turns out to be the following form
\begin{equation}
	\nu_{-N}=\Xi_N^{-1}\chi\nu_{-N+1}
	\label{eq:45}
\end{equation}
Using Eq.~(40)  in the second row of Eq.~(39), we get
\begin{equation}
	\nu_{-N+1}=-\Big(\frac{-\chi}{\Xi_{-N}}\eta+I_4\Xi_{-N+1}\Big)^{-1}\eta\nu_{-N+2}
	\label{eq:46}
\end{equation}
Using the following identity of matrix~\cite{Raicher13},
\begin{equation}
	(\delta_{1}I_4+\delta_2 I_4)^{-1}=\frac{1}{\delta_1}I_4-\frac{\delta_2}{\delta_1(\delta_1+\delta_2)}\mathcal{M}_1
	\label{eq:47}
\end{equation}
where $\delta_1$ and $\delta_2$ are scalar constants with $ \mathcal{M}_1$  defined in Eq.~(29), we can write Eq.~(42) as
\begin{equation}
	\nu_{-N+1}=-\Big[\frac{1}{\Xi_{-N+1}}\Sigma\Big]\eta\nu_{-N+2}
	\label{eq:48}
\end{equation}
In writing Eq.~(45), we have used Eq.~(44) and the condition $\chi^2=\eta^2=0$.\\
The generalized form of Eq.~(49) is, therefore,
\begin{equation}
	\nu_{-i}=-\Big[\frac{1}{\Xi_{-i}}\Sigma\Big]\eta\nu_{-i+1}
	\label{eq:49}
\end{equation}
Now Eq.~(49) can be written in the following form for $N > 0$ as the Eq.~(49)  also satisfies for $\mathcal{M}_2$. Thus we have
\begin{equation}
	\nu_{i}=-\Big[\frac{1}{\Xi_i}\Sigma\Big]\eta\nu_{i-1}
	\label{eq:50}
\end{equation}
The implication of results $\chi^2=\eta^2=0$ shows that the matrix  Eq.~(44)  can be truncated  to a finite order matrix of size $3\times 3$  for the case  $N=i\neq 0$ and $i=\pm 1$.\\
Finally, we have the following matrix equation
\begin{equation}
	\begin{pmatrix}
		\hat{ \Sigma}\Xi_{-1}I_4&\chi&0\\
		\eta&\hat{\Sigma}\Xi_{0}I_4&\chi\\
		0&\chi& \hat{\Sigma}\Xi_{1}I_4
	\end{pmatrix}
	\begin{pmatrix}
		\nu_{-1}\\
		\nu_0\\
		\nu_1
	\end{pmatrix}
	=0
	\label{eq:51}
\end{equation}
Writing $\nu_{-1}$ and $\nu_1$ in terms of $\nu_0$ we get
\begin{equation}
	\Sigma\Big( \Xi_0I_4-\frac{1}{\Xi_{-1}}\chi\eta-\frac{1}{\Xi_1}\eta\chi\Big)\nu_0=0
	\label{eq:52}
\end{equation}
Using Eq.~(30) and Eq.~(31),  a nontrivial solution of Eq.~(53) is deduced if the determinant vanishes. Therefore
\begin{equation}
	\det
	\begin{bmatrix}
		\Sigma\Xi_0I_4+e^2a^2(\omega'^2+2\Sigma m_{\text ph}^2)\Big(\frac{\mathcal{M}_1}{\Xi_{-1}}+\frac{\mathcal{M}_2}{\Xi_1}\Big)
	\end{bmatrix}
	=0
	\label{eq:53}
\end{equation}
For the given structure of matrices, $\mathcal{M}_1$ and $\mathcal{M}_2$, solution of Eq.~(49) gives the value of b, and   this reduces to a set of  two  scalar solutions:
\begin{equation}
	\Sigma\Xi_0\Xi_1+e^2a^2(\omega'^2+\Sigma m_{e}^2+m_{\text ph}^2)=0 \label{eq:54}
\end{equation}
and
\begin{equation}
	\Sigma\Xi_0\Xi_{-1}+e^2a^2(\omega'^2+\Sigma m_{e}^2+m_{\text ph}^2)=0
	\label{eq:55}
\end{equation}
where we assumed that for lasers with intensity $10^{23}{W/cm^2}$,  $e^2a^2\sim  m^{*}{}^2(=\omega_p^2)$. \\ For N=0,  the contribution of the term  $\Xi_0 \sim 0$. Using this argument in Eq.~(37) and solving it for b, we get the solutions of Eq.~(55) and Eq.~(56) as,
\begin{equation}
	b=\frac{k'\cdot p}{m_{\text ph}{}^2}\Big[1-\sqrt{1- \Big(\Sigma\frac{ea}{k'\cdot p}\Big)^2(\omega'^2+\Sigma m_e^2)+\omega_p^2}\Big]
	\label{eq:56}
\end{equation}
Substituting Eq.~(57) for b in Eq.~(42), we get
\begin{equation}
	\Xi_n\approx -2nm^{*2}(k'\cdot p)\sqrt{\Big[1+\Big(\Sigma\frac{ea}{k'\cdot p}\Big)^2(\omega'^2+\omega_p^2+\Sigma m_{e})^2}\Big]+2m_e^2n^2
	\label{eq:57}
\end{equation}
Now using Eq.~(57) and Eq.~(58) with the condition that $ea\sim m^*$ for lasers of intensities $(\text I\approx 10^{23}\si{Wcm^{-2}})$,  the solution of the second order differential equation in magnetized medium reduced to the following form:
\begin{equation}
	\psi'=e^{-\iota(p-bk')\cdot x'}\Big(1-\frac{\chi}{\Xi_1}e^{\iota\phi'}-\frac{1}{\Xi_-1}\eta e^{-\iota\phi'}\Big)\nu_0
	\label{eq:58}
\end{equation}
where
\begin{equation}
	\Xi_1=-2(k'\cdot p)\sqrt{1+\Big(\Sigma\frac{ea}{k'.p}\Big)^2(\omega'^2+\Sigma m_{e}^2+\omega_p^2)}
	\label{eq:59}
\end{equation}
and
\begin{equation}
	\Xi_{-1}=2(k'\cdot p)\sqrt{1+\Big(\Sigma\frac{ea}{k'.p}\Big)^2(\omega'^2+\Sigma m_{e}^2+\omega_p^2)}
	\label{eq:60}
\end{equation}
i.e.~$\Xi_1=-\Xi_{-1}$.
Using Eq.~(60) and Eq.~(61) and using definition of $\chi$ and  $\eta$ in Eq.~(59),we get
the final form of the solution of the Dirac equation is as
\begin{eqnarray}
	\psi'=\frac{1}{\sqrt{2\varepsilon_0}}\Big[1-\frac{{k'}}{m_e}\Big(b-\frac{e^2a^2}{\Xi_1}-\frac{e}{\Xi_1}\Big(\hat{\Sigma} {k'}\cdot {{A}}+{p}\cdot{{A'}}-{k'}\cdot{\epsilon'}\Big)\\ \nonumber -\frac{e}{4m_e}\Big(1+\frac{2k\cdot'p}{\Xi_1}\Big){{A}}-\frac{\iota e}{2m_e\Xi_-1}(\omega'^2+m_{ph}^2+\Sigma m_{e}^2){A}\Big]
	\cdot e^{-\iota(p-bk')\cdot x'}
	\label{eq:61}
\end{eqnarray}
where the value of $\nu_0$ is determined considering the following assumptions:
For $m^*{}^2 \ne 0$, Eq.~ (44)  has left  with $4\times 4$ vector equations. It implies that the degeneracy of $\mathcal {} $ is 4 with a choice of 4 degrees of freedom; hence, the value of   $\nu_0$ can be taken arbitrarily. Further, for $k'\cdot x' \to \pm \infty,$ the solution given in Eq.~(25)  reduces to a free particle Dirac equation. Thus, the final solution of the Dirac equation takes the following form
\begin{equation}
	\nu_0\approx u_{p\sigma}
	\label{eq:62}
\end{equation}
Also for $m_{\text ph}\ne 0$, we have
\begin{equation}
	\Xi_1\approx\Big[(\Sigma ea)^2-(\omega_p^2+k'^2)-\Sigma m_e^2 -m^2_{\text ph }\Big]
	\label{eq:63}
\end{equation}
and Eq.~(57) reduces to the following form
\begin{equation}
	b\approx-\frac{(k'\cdot p)}{m_{\text ph}^{2}}\Bigg[1+\frac{( e a)^2}{2}\left(\frac{\Omega^2_{ce}}{\omega'^2}\right)\left({2\omega_{p}^2}+\Sigma{m_{e}}^2\right)\Bigg]
	\label{eq:64}
\end{equation}
where $\Sigma^2$ is considered 1 for photon field and $m_{\text ph}=\omega_p$.The results for $b$ and $\Xi_1$ show that in the absence of plasma, our results deduce to well-known Volkov solutions except for a factor $\left(\Omega^2_{ce}/\omega^2\right)^2(2\omega_{p}^2+\Sigma {m_{e}}^2)$ which indicates the dependence of an external magnetic field, plasma frequency,  modified laser frequency, helicity of polarized states of relativistic electron and effective mass of photon. The solution for b given in Eq.~(62) gives a new kind of  Dirac solution of a relativistic electron near the SF-QED regime. The solution given in Eq.~(60) clearly shows a significant deviation from the Volkov results obtained earlier by Beers ~\cite{Beers72}. The form of present solutions in the plasma medium is contrary to the  Volkov solutions at intensities $ 10^{18}\si{Wcm^{-2}}\le\text I\le 10^{23}\si{Wcm^{-2}}$.\\ These solutions of the Dirac equation are functions of the laser frequency, cyclotron frequency, and plasma frequency, along with including the renormalized mass of the electron. These solutions can be reproduced to Volkov solutions under the conditions $\omega_p=\sqrt{\Sigma}m^{*}$ and $\Omega_{ce}=\omega$.
\section{Effect of magnetic field on Volkov solutions of Dirac equation}\label{3}
From Eq.~(60),we have
\begin{equation}
	b\approx-\frac{(k'\cdot p)}{m_{\text ph}^{2}}\Big[1+\frac{(\Sigma e a)^2}{2}\left(\frac{\gamma B_{\text {ext}}q_m}{\omega'^2}\right)^2\left(2\omega_p^2+\Sigma {m_{e}}^2\right)\Big]
	\label{eq:65}
\end{equation}
where $q_m$ is  electron charge mass ratio  and $\gamma =\sqrt{1+	p^2}$ for positive value of $m_{\text ph}$. The practical value of b depends on the renormalized mass of the electron, plasma frequency, laser intensity, and external magnetic field.
However, from Eq.~(14) it is obvious that  $\omega'\to \infty$ at $\Omega_{ce} \to \omega$. It shows singularity in Volkov solutions of the Dirac equation. Thus, on increasing the external magnetic field to acquire an electron cyclotron frequency equal to or above the laser frequency, Dirac's equation for  electron fails to explain the electron dynamics in magnetiized plasma and only the theory of  SF-QED is applicable. It implies that the Volkov solutions no longer validate the Dirac state functions at a  magnetic field strength $(\ge 1MG)$. The result also justifies that for optical laser and dense magnetized plasma $(m_{\text ph}\sim \omega')$, it corresponds to a laser of intensities $10^{18}\le\sim 10^{23}\si{Wcm^{-2}}$ ~\cite{Poisson99}. It insinuates that for $\chi\gg 1$, the Volkov wave functions are not good solutions, as was reported previously.  
\section{Conclusions and Discussion}
This study provides groundbreaking solutions to the complex problem of describing relativistic electrons in a laser-driven magnetoplasma, where both the electron's motion and its surrounding environment have significant and interdependent effects on its behavior. In contrast to traditional Volkov solutions, which treat isolated electrons in vacuum, our approach accounts for the crucial influence of photon mass resulting from interactions with the plasma environment. This addition significantly alters the electron’s behavior, unveiling new insights into how electrons interact with light under extreme conditions. As we approach the Schwinger limit of $1.32\times 10^{18}V/m$ or electromagnetic fields, our framework offers critical insights into how these extreme fields dramatically modify electron behavior. The novel incorporation of the electron's spin and its effective mass under such conditions further enriches our understanding of light-matter interactions in dense plasmas.
Our model also opens new avenues for investigating laser-assisted phenomena in plasmas, where the spinor nature of electrons becomes increasingly important. In particular, a strong background magnetic field can significantly modify Dirac electron solutions, amplifying the strength of the electromagnetic field at intensities near $1.32\times 10^{18}$V/m. or electromagnetic fields, our framework offers critical insights into how these extreme fields dramatically modify electron behavior. The novel incorporation of the electron's spin and its effective mass under such conditions further enriches our understanding of light-matter interactions in dense plasmas.
By exploring ultra-intense laser fields in magnetized plasma, we have advanced the understanding of relativistic electron dynamics, particularly the role of massive photons. This breakthrough provides a more accurate description of electron radiation and energy emission in the presence of powerful laser fields, a key factor in the design of next-generation laser-plasma accelerators.
Our findings also represents a general solution to the Dirac Equation beyond Volkov solutions, elucidating the helicity dependence of plasma electrons and consequently the polarization states of field quanta. It further elaborates that in the framework of strong-field quantum electrodynamics, particularly under the influence of magnetic field, multiple distinct solutions to the Dirac equation may emerge. These solutions correspond to  massive photon  mass states mediating with different polarization states of relativistic electrons.The novel results presented here were not accounted for in prior research efforts \cite {Erez14} .
\section*{Author Contributions}
B.S. Sharma laid the theoretical foundation for this work by validating the strong field quantum electrodynamics (SF-QED) approach. The remaining authors collaborated equally on manuscript preparation and discussions.

\section*{Funding Information}
No funding was provided by any agency.
\section*{Data Availability Statement}
No data were generated or analyzed in this theoretical study.
\\\\
{\color{green}ORCID iD\\ B.S.Sharma  http
	s://orcid.org/0000-0002-2327-9396}\\\\

\end{document}